\documentclass[preprintnumbers,aps,prb,amsmath,amssymb,nofootinbib,eqsecnum,superscriptaddress,floatfix]{revtex4}
\usepackage[dvips]{color}
\usepackage{amsmath}
\usepackage{graphicx}
\usepackage{slashed}
\usepackage{bm}
\begin{document}

\preprint{BU 07-08, MIT-CTP 3910, Isaac Newton Inst. N107074}

\title{
\large 
\bf 
Electron fractionalization for two-dimensional Dirac fermions
      }

\author{Claudio Chamon}
\affiliation{
Physics Department, Boston University, Boston, MA 02215, USA
        }

\author{Chang-Yu Hou}
\affiliation{
Physics Department, Boston University, Boston, MA 02215, USA
        }

\author{Roman Jackiw} 
\affiliation{
Department of Physics, Massachusetts Institute of Technology,
Cambridge, MA 02139, USA
            }

\author{Christopher Mudry} 
\affiliation{
Condensed Matter Theory Group,
Paul Scherrer Institut, CH-5232 Villigen PSI, Switzerland
        }

\author{So-Young Pi}
\affiliation{
Physics Department, Boston University, Boston, MA 02215, USA
        }

\author{Gordon Semenoff}
\affiliation{
Department of Physics and Astronomy, University of British Columbia, Vancouver, BC, Canada V6T 1Z1
        }

\date{\today}

\begin{abstract}
Fermion-number fractionalization without breaking of time-reversal symmetry
was recently demonstrated for a field theory in $(2+1)$-dimensional space and time
that describes the couplings between massive Dirac fermions,
a complex-valued Higgs field carrying an axial gauge charge of 2,
and a $U(1)$ axial gauge field. Charge fractionalization occurs 
whenever the Higgs field either supports vortices by itself, 
or when these vortices are accompanied by half-vortices in the axial gauge field. 
The fractional charge is computed by three different techniques. 
A formula for the fractional charge is given
as a function of a parameter in the Dirac Hamiltonian that breaks
the spectral energy-reflection symmetry. In the presence of a
charge $\pm1$ vortex in the Higgs field only, the fractional charge
varies continuously and thus can take irrational values.
The simultaneous presence  of a half-vortex in the axial gauge field 
and a charge $\pm1$ vortex in the Higgs field re-rationalizes
the fractional charge to the value $1/2$.
\end{abstract}

\maketitle


\section{
Introduction
        }

The concept of fractional charge emerged from quantum field theory
in 1976 when Jackiw and Rebbi showed 
that Bose fields can
induce a fractional fermion number $1/2$ for the 
relativistic fermions to which they couple.%
~\cite{Jackiw:1976}
The proper conditions for this mechanism of fractionalization
are the following.
First, the very notion of a fractional charge demands
that the fermion number is a good quantum number.
Second, the Bose fields must trigger the spontaneous breaking
of a symmetry that opens up a gap in the single-particle
fermionic spectrum within the Born-Oppenheimer approximation.
Third, the Bose fields must support local topological defects that
nucleate single-particle fermionic bound states in the close vicinity
to the defects. Fourth, this many-body quantum state 
is a finite energy eigenstate. 

The first requirement rules out mean-field descriptions
of superconductors that can otherwise satisfy the remaining
requirements.%
~\cite{Caroli:1964,Jackiw:1981,Read:2000}
The last requirement implies that the fractionalization of
the fermionic charge is a long-distance and low-energy property
of the many-body system, while the second and third ones
insure a degree of robustness against local perturbations.
This, in turn, suggests that the lessons learned from the
quantum field theories in Ref.~\onlinecite{Jackiw:1976}
could apply more generally to microscopic models 
encountered in solid state physics, thereby opening the possibility
of a ``table-top'' measurement of the fractional charge.

In fact, the work of Su, Schrieffer, and Heeger implies that the
one-dimensional example of Ref.~\onlinecite{Jackiw:1976} can be
thought of as an effective field theory that captures the relevant
interactions between phonons and electrons in polyacetylene.%
~\cite{Su:1979,Jackiw:1981npb} Excitations with exotic quantum numbers
(in relation to the fundamental electron constituents of the system),
such as neutral objects carrying spin 1/2 or charge $\pm 1$ objects
carrying zero spin, localize around a domain wall in the dimerization
pattern of polyacetylene at the cost of a finite energy. Subsequent to
this work, it was shown that exotic fermionic quantum numbers in
one-dimension are not restricted to fractional values,%
\cite{Goldstone:1981,Rice:1982,Jackiw:1983,Kivelson:1983} but can be
tuned continuously by a small breaking of an energy-reflection
symmetry assumed in Refs.~\onlinecite{Jackiw:1976}
and~\onlinecite{Su:1979}, and defined below.

With the discovery of the fractional quantum Hall effect,
a different paradigm for charge fractionalization,
one in which spontaneous symmetry breaking plays no role,
was proposed by Laughlin
for two-dimensional systems with strong breaking
of time-reversal symmetry.\cite{Laughlin:1983,Halperin:1984}
Central to this paradigm is the notion of topological order, 
a global property that characterizes an otherwise featureless 
incompressible liquid state of matter by the finite
degeneracy of the ground state if the system is defined 
on a surface of non-trivial topology, with the degeneracy depending on
the genus of the surface.~\cite{Wen:1990} 
The fractional charge is intimately connected to 
this ground-state degeneracy,
leaving no room for a continuoulsy varying fractional charge
and, in particular, for an irrational charge.
Since then, the preferred route towards charge fractionalization
without time-reversal symmetry in two and higher dimensions 
has occulted any mechanism based on 
sponteneous symmetry breaking, presumably because
it is believed that the energy cost for fractional
charges is prohibitive in all but one dimension.

However, as a matter of principle,
this need not be so as was already shown by Jackiw and 
Rebbi in three-dimensional space when coupling Yang-Mills
fields through the minimal coupling to Higgs
fields and to Dirac fermions.
Quantization of the Dirac fermions in the static
background of a t'Hooft-Polyakov monopole
nucleates a fermionic bound states with the fractional
charge $1/2$ at a finite cost in energy.\cite{Jackiw:1976}

Of course, one might object that this three-dimensional 
example of charge fractionalization
is unlikely to be realized on the energy scale of
the electron volt that governs solid state physics,
a prerequisite for a table-top measurement of charge
fractionalization.%
~\cite{footnote on cold atoms} 
We do not know of a realistic
three-dimensional model for band electrons coupled 
to bosonic collective modes that mimics Dirac fermions 
and Higgs fields coupled with each others and coupled
minimally to Yang-Mills in the continuum limit.
In two and three dimensions band-theory generically predicts
an insulating or a metallic state of matter. 
In one-dimension the Fermi surface is generically realized 
by an even number of discrete points, 
thus providing the low-energy and long-wave-length limit 
of the tight-binding model with a Dirac structure for free.

Semimetals, the most famous example of which is graphite, are
exceptions to the hegemony of the band-insulating and of the metallic
states of matter. Graphite is made of sheets of graphene, a honeycomb
lattice made of carbon ions bound through $sp^2$ orbitals, and where
the fourth valence electron of each atom lazily revels predominantly
between planar nearest-neighbor sites. The Fermi surface at
half-filling for an isolated graphene sheet is made of two isolated
points.\cite{Wallace:1947} The excitation spectrum around these two
Fermi points endows the band electrons with a four-component Dirac
structure owing to the Nielsen-Ninomiya theorem.~\cite{Nielsen:1981}

\begin{figure}
\includegraphics[angle=0,scale=0.4]{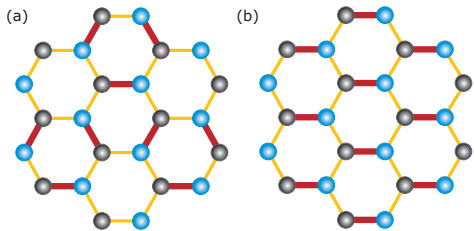}
\caption{ (color online).  
  The honeycomb lattice is shown in (a) and (b). The honeycomb lattice
  has 2 interpenetrating Bravais sublattices colored in blue
  and black, respectively. 
  The electronic hopping amplitude is enhanced on thick red bonds while it
  is reduced on the thin yellow bonds relative to the magnitude $t$ of
  the nearest-neighbor hopping amplitude. The so-called
  K\'ekule pattern dimerization pattern in (a) 
  opens an energy gap for the single-particle fermionic levels
  and it maps to the complex-valued Higgs field of Ref.~\onlinecite{Hou:2007} 
  in the continuum limit. A dimerization pattern that shifts the relative 
  separation of the Dirac points is shown is shown in (b). In the continuum limit,
  it maps to the axial vector potential introduced in Ref.~\onlinecite{Jackiw:2007}.  
         }
\label{fig:honeycomb dimerization}
\end{figure}

Although this example of fermion-doubling is often viewed
as a curse for the realization of quantum anomalies,%
\cite{Semenoff:1984,Haldane:1988,Ryu:2007}
it is this very property that opens the door to
charge fractionalization without 
the breaking of time-reversal symmetry
through spontaneous symmetry breaking, as shown by 
Hou, Chamon, and Mudry.~\cite{Hou:2007}
The real-valued static fluctuations depicted in 
Fig.~\ref{fig:honeycomb dimerization}(a)
about the uniform nearest-neighbor
hopping amplitudes of graphene are, in the continuum limit,
represented by a complex-valued Higgs field
that interacts with the four-component Dirac fermions.
In the Born-Oppenheimer approximation, 
a constant value of this complex-valued Higgs field
breaks spontaneously an effective axial $U(1)$ symmetry
of the continuum limit and opens up a gap 
in the single-particle fermion spectrum.
If the phase of this complex-valued Higgs field
is defective in that it carries a vortex,
it nucleates single-particle mid-gap states
that carry the fractional charge $\pm 1/2$ per state.
The energy cost is not finite, however.
In the continuum approximation, it grows logarithmically
with the separation between the vortices.
On the lattice, it even grows linearly with the vortex 
separation if the wave vector of the fluctuation
of the hopping amplitude is commensurate with the 
reciprocal lattice.

Jackiw and Pi showed that 
the energy cost of a vortex in the complex-valued Higgs
field can be made finite if the complex-valued
Higgs field and the Dirac fermions couple minimally to
two real-valued Bose fields that realize
the vector components of an axial gauge field
and if this axial vector gauge field also
supports vortices.%
\cite{Jackiw:2007}  
A honeycomb-lattice regularization of an
axial gauge field without a vortex
is shown in Fig.~\ref{fig:honeycomb dimerization}(b).%
~\cite{Chamon:2007}
Alternative realizations of an axial vector potential
also arise when the graphene sheet is curved%
\cite{Morozov:2006,Morpurgo:2006,Cortijo:2007}
or
wrapped into fullerenes,%
~\cite{González:1992,Pachos:2007}
into nanotubes,%
~\cite{Kane:1997,Chamon:2000}
and about a cone.%
~\cite{Lammert:2000,Osipov:2001}

Charge fractionalization in one dimension can be tuned continuously 
by breaking an energy-reflection symmetry, which is defined below. 
This property survives in two dimensions and gives a mechanism for 
charge fractionalization that is fundamentally different 
(and thus potentially observable)
from the mechanism for charge 
fractionalization that relies on topological 
order. By relaxing the condition of a finite energy
to that of a logarithmically diverging energy,
it was argued in Ref.~\onlinecite{Hou:2007} that
a small staggered chemical potential that distinguishes
carbon ions sitting on the now nonequivalent triangular
sublattices of the honeycomb lattice
can make the fractional charge irrational.
This irrational charge was calculated analytically
in the continuum limit and numerically for a lattice
regularization in Ref.~\onlinecite{Chamon:2007}.
Remarkably, it was also found that the condition for finite energy in the continuum limit, i.e., the
presence of a vortex in the axial vector potential,
removed any dependence of the fractional charge on
the staggered chemical potential. 

The purpose of this paper is to give three detailed 
and alternative derivations of the fractional charge 
that supplement the derivation from 
Ref.~\onlinecite{Chamon:2007}.
Graphene is of course not the only road to a semimetal in
two dimensions. Threading the elementary plaquettes of 
a square lattice with half a flux quantum%
~\cite{Lieb:1994}
also realizes two nonequivalent Dirac points 
at half-filling. The Higgs field is then realized by
a columnar pattern of dimerization whereas the axial
vector gauge field is realized by a staggered pattern of
dimerization.\cite{Chamon:2007} It was shown numerically 
in Ref.~\onlinecite{Seradjeh:2007}
that the $\mathbb{Z}^{\ }_{4}$ vortex
defined by the four possible columnar patterns occupying
the four quadrants of the square lattice pins the fractional
charge $\pm1/2$ at the site where the four columnar patterns meet.
The $\mathbb{Z}^{\ }_{4}$ vortex is a discontinuous version
of the vortices studied numerically in 
Ref.~\onlinecite{Chamon:2007}.
The fact that different lattice regularization of vortices
carry the same fractional charge illustrates the
fact that the fractional charge is independent of the 
short-distance regularization.
This property will become obvious in the analytical 
calculations  of the fractional charge that we are going to present.

The paper is organized as follows.
The quantum field theory is defined in Section~\ref{sec: Definitions}.
The charge induced by the vortices in the Higgs or axial gauge
fields is related to the spectral asymmetry in 
Section~\ref{sec: Quantum mechanical analysis}.
The spectral asymmetry is computed in 
Setion~\ref{sec: Spectral asymmetry and fractional charge}.
The fractional charge is computed a second time by 
a perturbative expansion of the Dirac propagator in
Section~\ref{sec: Field theory analysis}.
Finally, the fractional charge is computed on the basis
of symmetry arguments in
Section~\ref{sec: Induced charge from symmetry arguments}. A summary comprises the last Section~\ref{sec:summary}

\section{
Definitions
        }
\label{sec: Definitions}

In recent Letters,\cite{Hou:2007,Jackiw:2007,Chamon:2007} 
we have developed the theory of charge
fractionalization in planar models with topological defects
encoded by vortices.
The models are elaborations on graphene, with dynamics
linearized around two Dirac points (the two inequivalent
points in the first Brillouin zone at which the conduction
and valence bonds of graphene meet).
In a familiar fashion, the Schr\"odinger
equation for the Bloch states
at low energies and long wave length
measured relative to the Dirac points
takes a Dirac-like $4 \times 4$ matrix form, for a four-component
``spinor'', which interacts with further scalar and gauge fields.
The scalar field and the gauge fields are induced
by fluctuations in the hopping amplitudes of the underlying
microscopic tight-binding model.
In this Section, we start with definitions.

In second quantization, the planar Hamilton density reads 
\begin{equation}
\mathcal{H}=
\psi^{\dag} 
\big[
\boldsymbol{\alpha} 
\cdot 
\left(\boldsymbol{p} 
- 
\gamma^{\ }_5 \boldsymbol{A}^{\ }_{5} \right) 
+ 
\beta 
\left( 
\varphi^{\ }_{1}
- 
\mathrm{i} 
\gamma^{\ }_5  
\varphi^{\ }_{2}  
\right) 
+
R\,\mu
\big]
\psi
\equiv
\psi^{\dag}\, H\,\psi.
\label{eq:Hamiltonian}
\end{equation}
Here, 
$\psi^{\dag}$ and $\psi$ are creation and annihilation 
operators for four-components Dirac fermions, respectively,
$\boldsymbol{p}= -\mathrm{i} (\partial^{\ }_x, \partial^{\ }_y )$; 
$\boldsymbol{A}^{\ }_{5}$ 
is an axial vector gauge potential $(A^x_5,A^y_5)$; 
$\varphi^{\ }_{1}$ and $\varphi^{\ }_{2}$ 
are the real and imaginary parts of a complex scalar field 
$\varphi= \varphi^{\ }_{1}+\mathrm{i} \varphi^{\ }_{2}$; 
and $\mu$ is a field that acts like a staggered
chemical potential -- the staggering is governed by the matrix $R$. 
All fields depend on the three-vector $x^\mu=(t,\boldsymbol{r})=(t,x,y)$.
The matrices in $H$ are conventional $4 \times 4$ Dirac matrices:
\begin{subequations}
\begin{equation}
\boldsymbol{\alpha}=(\alpha^x, \alpha^y) \equiv 
\left(\begin{array}{cc}
\boldsymbol{\sigma} & 0 \\0 & -\boldsymbol{\sigma} 
\end{array}\right), 
\qquad 
\beta\equiv 
\left(\begin{array}{cc} 0 & 1 \\ 1 & 0 
\end{array}\right), 
\qquad 
\gamma^{\ }_5 \equiv -\mathrm{i} \alpha^x \alpha^y \alpha^z= 
\left(\begin{array}{cc} 
1& 0 \\ 0& -1 
\end{array}\right),
\end{equation}  
where the ``third'' $\alpha$-matrix,
\begin{equation}
\alpha^z \equiv  
\left(\begin{array}{cc}
\sigma^{\ }_{3} & 0 \\0 & -\sigma^{\ }_{3} 
\end{array}\right),
\end{equation}
\end{subequations}
participates in the definition of $\gamma^{\ }_5$ and also coincides with
the matrix $R\equiv \alpha^z$.
(The matrices $\sigma^{\ }_{1,2,3}$ are the standard Pauli matrices.)

The Lagrange density corresponding to Eq.~(\ref{eq:Hamiltonian}),
\begin{subequations}
\begin{equation}
\mathcal{L} = 
\mathrm{i} \psi^{\dag} \partial^{\ }_t \psi 
- 
\mathcal{H},
\end{equation} 
is presented in covariant notation as
\begin{equation}
\label{eq:Lagrangian}
\mathcal{L}= 
\bar{\psi} 
\big[
\gamma^\nu 
\left(\mathrm{i} \partial^{\ }_\nu + \gamma^{\ }_5 A^{\ }_{5\nu}\right) 
- 
\left( \varphi^{\ }_{1} -\mathrm{i} \gamma^{\ }_5 \varphi^{\ }_{2} \right) 
-
\gamma^3 \mu
\big] 
\psi ,
\end{equation}
with
\begin{equation}
\gamma^0\equiv \beta, \qquad  \bar{\psi} \equiv \psi^\dag \gamma^0, \qquad
\boldsymbol{\gamma} \equiv \beta \boldsymbol{\alpha} =
\left(\begin{array}{cc} 
0 & -\boldsymbol{\sigma} \\ \boldsymbol{\sigma} & 0
\end{array}\right),
\qquad 
\gamma^3\equiv \beta R =\beta \alpha^z= 
\left(\begin{array}{cc} 
0 & -\sigma^{\ }_{3} \\ \sigma^{\ }_{3} & 0
\end{array}\right). 
\end{equation}
\end{subequations}
[In Eq.~(\ref{eq:Hamiltonian}), we set 
the axial scalar gauge potential $A^{0}_5$ to zero.] 

Two gauge transformations leave the model unchanged. 
There is the local axial gauge symmetry
\begin{equation}
\psi \to e^{\mathrm{i} \omega \gamma^{\ }_5} \psi, 
\qquad 
A^{\ }_{5\nu} \to  A^{\ }_{5\nu} + \partial^{\ }_{\nu} \omega, 
\qquad 
\varphi \to e^{2 \mathrm{i} \omega} \varphi, 
\qquad 
\mu \to \mu,
\end{equation}
where $\omega$ is a real-valued field and the index $\nu=t,x,y$.
Also, there is the global phase symmetry, 
\begin{equation}
\psi \to e^{\mathrm{i} \varpi} \psi, 
\qquad A^{\ }_{5\nu} \to  A^{\ }_{5\nu}
\qquad \varphi \to \varphi, 
\qquad \mu \to \mu,
\end{equation}
where $\varpi$ is a real-valued number,
which acts on the four components of the spinors. 
The latter leads to the conserved fermion (charge) number current.
\begin{equation}
J^{\nu} \equiv 
\bar{\psi} 
\gamma^{\nu}
\psi=
(\rho,\boldsymbol{j} )= 
(\psi^{\dag} \psi, \psi^\dag \boldsymbol{\alpha} \psi),
\qquad
\partial^{\ }_{\nu} J^{\nu}=0.
\end{equation}
We shall show that the charge
\begin{equation}
Q=\int d^2 r\; \rho(\boldsymbol{r}) 
\end{equation}
fractionalizes when the background bose fields are topologically nontrivial.

The model possesses the usual discrete symmetries 
under the parity transformation $\mathcal{P}$ defined by 
\begin{equation}
\label{eq:parity-trans}
\mathcal{P}:
\left\{
\begin{array}{ll}
(t,x,y
)& 
\to (t,-x,y),
\\
\psi(t,x,y) 
& 
\to \mathrm{i} \gamma^3 \gamma_1 \psi(t,-x,y),
\\
A^{t,y}_5(t,x,y) 
& 
\to A^{t,y}_5(t,-x,y),
\\
A^x_5(t,x,y) 
& 
\to -A^x_5(t,-x,y),
\\
\varphi(t,x,y) 
& 
\to \varphi(t,-x,y),
\\
\mu(t,x,y) 
&
\to -\mu(t,-x,y),
\end{array}
\right.
\end{equation}
the charge conjugate transformation $\mathcal{C}$ defined by
\begin{equation}
\mathcal{C}:
\left\{
\begin{array}{ll}
\psi^{\ }_i
& 
\to \gamma^1_{ij} \bar{\psi}^{\ }_j,
\\
A^{\ }_{5\nu} 
& 
\to -A^{\ }_{5\nu},
\\
\varphi 
& 
\to \varphi^*,
\\
\mu 
&
\to \mu,
\end{array}
\right. 
\end{equation}
and the time-reversal transformation $\mathcal{T}$ defined by
\begin{equation}
\mathcal{T}:
\left\{
\begin{array}{ll}
(t,x,y)& \to (-t,x,y),
\\
\psi(t,x,y) & \to \gamma^1 \gamma^5 \psi^{\dag}(-t,x,y),
\\
A^{\nu}_5(t,x,y) & \to A^{\nu}_5(-t,x,y),
\\
\varphi(t,x,y) & \to \varphi^{*}(-t,x,y),
\\
\mu(t,x,y) &\to \mu(-t,x,y),
\end{array}
\right.
\end{equation}
where one should remember that $\mathcal{T}$ is antiunitary so that
complex conjugation of coefficients is implied.

The theory possesses another discrete
symmetry
\begin{equation}
\label{eq:Additional-transformation}
\left\{
\begin{array}{ll}
\psi 
& 
\to \mathrm{i} \gamma^3 \gamma^5 \psi,
\\
A^{\ }_{5\nu} 
& 
\to -A^{\ }_{5\nu},
\\
\varphi 
& 
\to \varphi^*,
\\
\mu & \to -\mu.
\end{array}
\right.
\end{equation}
In the lattice (the honeycomb lattice relevant to graphene, for
example), the definition of parity depends on the axis used for the
reflection; the transformation in Eq.~(\ref{eq:parity-trans})
corresponds to a reflection with respect to an axis that cuts through
the bonds of the honeycomb lattice.

When the staggered chemical potential $\mu$ is dropped, i.e., the last term in the square brackets of
Eq.~(\ref{eq:Hamiltonian}) or Eq.~(\ref{eq:Lagrangian}) is absent,
the matrix $R$ anticommutes with the remaining matrices in the
single-particle Hamiltonian $H$ of Eq.~(\ref{eq:Hamiltonian}). 
Therefore, $R$ maps positive energy eigenfunctions $\Psi^{\ }_{E}$ 
to negative energy eigenfunctions $\Psi^{\ }_{-E}$
and vice-versa,
\begin{equation}
H\big|^{\ }_{\mu=0} \Psi^{\ }_E=
E \Psi^{\ }_E,
\qquad 
R\Psi^{\ }_E= 
\Psi^{\ }_{-E}. 
\end{equation}
We call this an ``energy-reflection symmetry''. 

We shall examine the Dirac theory with a specific vortex
configuration for the Bose field $\varphi$, 
taken as a static background,
and with another specific vortex
configuration for the axial gauge field $A^{\nu}_{5}$, 
also taken as a static background.
The polar decomposition of the scalar field $\varphi$ is
\begin{equation}
\label{eq:scalar-field-profile}
\varphi(\boldsymbol{r}) =
\phi(r)\, 
e^{\mathrm{i} n \theta},
\qquad
r=\sqrt{x^2+y^2},
\qquad
\theta=\mathrm{arctan}\,\frac{y}{x},
\end{equation}
where the magnitude 
$\phi$ of $\varphi$
vanishes at the origin 
$\phi(r=0)$
and tends to a nonvanishing 
$\phi(\infty)$ 
for large $r$. The integer $n$ measures the vorticity
encoded by the singular nature of the phase of the complex
field $\varphi$ at the origin.
The axial gauge potential vanishes in the time component
\begin{subequations}
\label{eq:gauge-field-profile}
\begin{equation}
\label{eq:scalar-gauge-field-profile}
A^{0}_5(\boldsymbol{r}) =0, 
\end{equation}
while the spatial component reads
\begin{equation}
\label{eq:vector-gauge-field-profile}
A^{i}_5(\boldsymbol{r}) = 
- n \epsilon^{\ }_{ij} \frac{r^j}{r^2} a^{\ }_{5}(r) , 
\end{equation}
\end{subequations}
where $a^{\ }_{5}(r)$ 
vanishes at the origin and tends to $1/2$
at large $r$. 
The line integral over Eq.~(\ref{eq:vector-gauge-field-profile})
along any closed curve that encircles once the origin
yields the same number, proportional to the vorticity $n$.
Finally, the chemical potential $\mu$,
also taken as a static background,
is without topological structure and
achieves a nonvanishing value $\mu(\infty)$ at infinity. We shall take 
$\mu$ to depend only on $r$, 
but it could also be constant.

In the absence of the staggered chemical potential, the Dirac equation
possesses $|n|$ zero-energy, normalizable solutions. These are the
mid-gap states, eigenstates of $R$. Mostly, we consider the $n=-1$
case, with a single mid-gap state, $\Psi^{\ }_0$, which remains bound even
in the presence of the axial vector potential; 
turning on the axial vector potential changes the wave function profile, 
but the zero eigenvalue remains. 
We assume that there are no other bound states.
When the staggered chemical potential is present, 
but never very large, 
the mid-gap state migrates to a shifted eigenvalue; 
however it still remains isolated in the gap.

\section{
Quantum mechanical analysis
        }
\label{sec: Quantum mechanical analysis}

The following argument shows that without the staggered chemical potential $\mu$
the charge is $-1/2$ when there is a single normalizable mid-gap state
$\Psi^{\ }_{0}$ that is unoccupied.
(When this mid-gap state is occupied, the charge is $-1/2+1=+1/2$.)
The charge density arises from filling the 
negative energy continuum states 
of the Dirac equation,
\begin{eqnarray}
\rho(\boldsymbol{r})&=&
\int\limits_{-\infty}^0 dE\, 
\Big[ 
\Psi^{\dag}_E(\boldsymbol{r}) \Psi^{\ }_E(\boldsymbol{r})
- 
\Upsilon^{\dag}_E(\boldsymbol{r})\Upsilon^{\ }_E(\boldsymbol{r}) 
\Big] 
\nonumber\\
&=& 
\frac{1}{2} 
\int\limits_{-\infty}^\infty dE\, 
\Big[ 
\Psi^{\dag}_E(\boldsymbol{r})\Psi^{\ }_E(\boldsymbol{r})
- 
\Upsilon^{\dag}_E(\boldsymbol{r})\Upsilon^{\ }_E(\boldsymbol{r}) 
\Big],
\label{eq:continuum-substraction} 
\end{eqnarray}
where the second equality follows from the first due to the energy-reflection symmetry present in the problem at $\mu=0$. The quantity
$\Upsilon^{\dag}_E\Upsilon^{\ }_E$ 
is constructed from reference states which solve a
Dirac equation with a topologically trivial background and also
possess the energy-reflection symmetry. In other words, the
topologically determined charges that we compute are measured relative
to a reference charge of a system with a topologically trivial
background and possessing the energy-reflection symmetry. 
This procedure is needed to remove infinities. The
reference wave functions $\Upsilon^{\ }_E$ form a complete set. The continuum
wave functions $\Psi^{\ }_E$ in the presence of the vortex -- we call
them the vortex states -- are not complete; the mid-gap state is
missing
\begin{equation}
\delta(\boldsymbol{r}-\boldsymbol{r}')=
\int\limits_{-\infty}^\infty dE\, 
\Upsilon^{\dag}_E(\boldsymbol{r})\Upsilon^{\ }_E(\boldsymbol{r}')= 
\int\limits_{-\infty}^\infty dE\, 
\Psi^{\dag}_E(\boldsymbol{r})\Psi^{\ }_E(\boldsymbol{r}')
+
\Psi^{\dag}_0(\boldsymbol{r})\Psi^{\ }_0(\boldsymbol{r}').
\label{eq:completeness I}
\end{equation}
It therefore follows from combining
Eq.~(\ref{eq:continuum-substraction})
with 
Eq.~(\ref{eq:completeness I})
that 
\begin{equation}
\rho(\boldsymbol{r})=
-
\frac{1}{2}
\Psi^{\dag}_0(\boldsymbol{r})\, \Psi^{\ }_0(\boldsymbol{r})
\label{eq:density-with-SLS}
\end{equation}
and
\begin{equation}
Q= \int d^2 r \;\rho(\boldsymbol{r})=-
\frac{1}{2} .
\end{equation}

In the presence of the staggered chemical potential $\mu$, the
energy-reflection symmetry is no longer available to pass from the
first to the second equality of
Eq.~(\ref{eq:continuum-substraction}). However, we may proceed as
follows. We suppose that the reference states still possess the energy-reflection symmetry, 
so in Eq.~(\ref{eq:continuum-substraction})
 we may still use this symmetry for them.~\cite{footnote-on-spectral-sym}
\begin{equation}
\int\limits_{-\infty}^0 dE\,
\Upsilon^{\dag}_E\Upsilon^{\ }_E =
\frac{1}{2} 
\int\limits_{-\infty}^{\infty} dE\, 
\Upsilon^{\dag}_E\Upsilon^{\ }_E
=
\frac{1}{2} 
\int\limits_{-\infty}^{\infty} dE\, 
\Psi^{\dag}_E\Psi^{\ }_E
+
\frac{1}{2} 
\Psi^{\dag}_b\Psi^{\ }_b
\label{eq:completeness II}
\end{equation}
The last equality is again the statement of completeness of the continuum
reference states and the continuum vortex states supplemented by the
isolated bound state $\Psi^{\ }_b$, which is no longer at zero energy but
has migrated to some other value in the gap between the continuum
states. Using Eq.~(\ref{eq:completeness II}) in the first equality of
Eq.~(\ref{eq:continuum-substraction}) leaves
\begin{equation}
\rho(\boldsymbol{r})= 
-
\frac{1}{2} 
\Psi^{\dag}_b(\boldsymbol{r})\Psi^{\ }_b(\boldsymbol{r}) 
-
\frac{1}{2} 
\int\limits_{-\infty}^{\infty} dE\, 
\mathrm{sign}(E)\,
\Psi^{\dag}_E(\boldsymbol{r}) \Psi^{\ }_E(\boldsymbol{r}). 
\label{eq: final result sec QM}
\end{equation}
It remains to evaluate the remaining integral, which is recognized as
the ``$\eta$-invariant'', also called ``spectral asymmetry''.
Note that with energy-reflection symmetry the integral vanishes, 
leading to the previous result~(\ref{eq:density-with-SLS}). 
In the above derivations, it is assumed that the ``vacuum''  is defined 
with the mid gap state $\Psi^{\ }_0$ unoccupied, 
and furthermore that the migrated state $\Phi^{\ }_b$ has positive energy 
so that it remains unoccupied in the definition of the vacuum. 
If the mid-gap state is occupied and/or the migrated state has negative energy, 
there occurs a sign change in Eq.~(\ref{eq:density-with-SLS}) 
that affects to the first term of Eq.~(\ref{eq: final result sec QM}).

\section{
Spectral asymmetry and fractional charge
        }
\label{sec: Spectral asymmetry and fractional charge}

We begin by putting the Hamiltonian Eq.~(\ref{eq:Hamiltonian}) in a more convenient form. This is done with the following unitary transformation
\begin{subequations}
\begin{equation}
H\to \hat{H}=T H T^{-1},
\end{equation}
where
\begin{equation}
T \equiv \begin{pmatrix} 
\mathrm{i} \sigma_{-}
& 
\sigma_{+}
\\ 
-\mathrm{i} \sigma_{+}
&
\sigma_{-}  
\end{pmatrix}, \qquad 
T^{-1}=T^{\dag}, \qquad \sigma_{\pm}\equiv \frac{1}{2} (1\pm \sigma_3). 
\end{equation}
\end{subequations}
The result for $\hat{H}$ is 
\begin{subequations}
\begin{equation}
\hat{H}
=
\begin{pmatrix} 
-\mu 
& 
\mathcal{D}
\\ 
\mathcal{D}^{\dag} 
&
\mu  
\end{pmatrix}
\end{equation} 
where the differential operator $\mathcal{D}$ and its adjoint $\mathcal{D}^\dag$ are given by
\begin{eqnarray}
\label{eq:D}
\mathcal{D}&=&
\mathrm{i} \sigma^i( \boldsymbol{\partial}_i+\epsilon_{ij} A_5^j) 
+
\mathrm{i} \varphi^{\ }_{1}
+
\sigma^{\ }_{3} \varphi^{\ }_{2},
\\ 
\label{eq:Ddagger} 
\mathcal{D}^\dag&=&
\mathrm{i} \sigma^i( \boldsymbol{\partial}_i-\epsilon_{ij} A_5^j) 
-
\mathrm{i} \varphi^{\ }_{1}
+
\sigma^{\ }_{3} \varphi^{\ }_{2}.
\end{eqnarray}
\end{subequations}
With the factorization of the time dependence
$\left(u,v\right)=e^{-\mathrm{i}Et}\left(u^{\ }_E,v^{\ }_E\right)$, 
the stationary Dirac equation reads
\begin{subequations}
\begin{equation}
\begin{pmatrix} 
-\mu  
& 
\mathcal{D} 
\\ 
\mathcal{D}^\dag
& 
\mu
\end{pmatrix} 
\begin{pmatrix}
u^{\ }_E
\\
v^{\ }_E
\end{pmatrix}= 
E 
\begin{pmatrix}
u^{\ }_E
\\
v^{\ }_E
\end{pmatrix}
\label{diracham}
\end{equation}
or in terms of components,
\begin{eqnarray}
\label{eq:u}
\mathcal{D}v^{\ }_E = \left(E+ \mu \right)u^{\ }_E,
\\
\label{eq:v} 
\mathcal{D}^\dag u^{\ }_E = \left(E - \mu \right)v^{\ }_E.
\end{eqnarray}
\end{subequations}
Generally, $\mu$ can be a function on space-time.  In the
remainder of this Section, we shall set it to be a positive constant; 
the other background fields, $\varphi(\boldsymbol{r})$ 
and $\boldsymbol{A}^{\ }_{5}(\boldsymbol{r})$ 
are position dependent and static, with asymptotes quoted in 
Eqs.~(\ref{eq:gauge-field-profile}) and~(\ref{eq:scalar-field-profile}).

\subsection{Zero-mode solutions}

It is convenient to begin by considering two special cases, particular solutions of Eqs.~(\ref{eq:u})
and (\ref{eq:v}) where either $E=\mu $ or $E=-\mu$. These solutions
would become zero modes of the Hamiltonian when $\mu=0$ and they play a special role even when $\mu \neq 0$.  

Let us begin with the case where the energy eigenvalue $E=\mu$. 
Then, from Eqs.~(\ref{eq:u}) and (\ref{eq:v}), 
it follows that
\begin{subequations}
\begin{eqnarray}
\label{eq:ua} 
&&
\mathcal{D}^\dag  u^{\ }_\mu(\boldsymbol{r})=0,
\\
\label{eq:va} 
&&
u^{\ }_\mu(\boldsymbol{r})=
\frac{1}{2\mu }\mathcal{D}
v^{\ }_\mu(\boldsymbol{r}).
\end{eqnarray} 
\end{subequations}
First we observe that if $v^{\ }_\mu(\boldsymbol{r})$ were identically zero, 
$u^{\ }_\mu(\boldsymbol{r})$ would also vanish and there is no solution.
So we assume that $v^{\ }_\mu(\boldsymbol{r})\neq 0$. 
Then, operating with $\mathcal{D}^\dag$ on Eq.~(\ref{eq:va}) and using
Eq.~(\ref{eq:ua}) yields 
$\mathcal{D}^\dag \mathcal{D} v^{\ }_\mu(\boldsymbol{r})=0$.
The following argument implies that $\mathcal{D} v^{\ }_\mu(\boldsymbol{r})=0$.  Consider
\begin{equation}
\label{eq:argument}
0=
\int d^2r~ 
v^{\dag}_{\mu}(\boldsymbol{r})
\mathcal{D}^\dag \mathcal{D} v^{\ }_\mu(\boldsymbol{r}) = 
\int d^2r~\vert \mathcal{D} v^{\ }_\mu(\boldsymbol{r}) \vert^2.
\end{equation} 
Here, we are assuming that the spinor $v^{\ }_\mu(\boldsymbol{r})$ 
obeys boundary conditions so that the differential operator 
$\mathcal{D}^\dag$ is indeed the adjoint of $\mathcal{D}$, 
i.e., surface terms produced by partial integrations in the intermediate steps in (\ref{eq:argument}) vanish. Since the last integral vanishes, 
its positive semi-definite integrand must also vanish 
and we conclude that 
\begin{equation}
\label{eq:vb}
\mathcal{D} v^{\ }_\mu(\boldsymbol{r})=0
\end{equation} 
while Eq.(\ref{eq:ua}) implies that $u^{\ }_\mu(\boldsymbol{r})=0$.

Thus, we find that, when Eq.(\ref{eq:vb}) possesses a normalizable solution, 
there exists a positive energy bound state with $E=\mu$,
\begin{equation}
\begin{pmatrix} 
-\mu  
& 
\mathcal{D} 
\\ 
\mathcal{D}^\dag 
& 
\mu
\end{pmatrix} 
\begin{pmatrix}
0
\\
v^{\ }_\mu(\boldsymbol{r}) 
\end{pmatrix}= 
\mu  
\begin{pmatrix}
0
\\
v^{\ }_\mu(\boldsymbol{r}) 
\end{pmatrix},
\qquad
\int d^2r |v^{\ }_\mu(\boldsymbol{r})|^2=1.
\end{equation}
Similar reasoning establishes the occurrence of a negative energy
bound state with $E=-\mu$ when there exists a normalizable solution
of the equation $ \mathcal{D}^\dag u^{\ }_{-\mu}(\boldsymbol{r})=0$,
\begin{equation}
\begin{pmatrix} 
-\mu  
& 
\mathcal{D} 
\\ 
\mathcal{D}^\dag 
& 
\mu
\end{pmatrix} 
\begin{pmatrix}
u^{\ }_{-\mu}(\boldsymbol{r})
\\ 
0
\end{pmatrix} = 
-\mu 
\begin{pmatrix}
u^{\ }_{-\mu}(\boldsymbol{r})
\\ 
0
\end{pmatrix},
\qquad
\int d^2r |u^{\ }_{-\mu}(\boldsymbol{r})|^2=1.
\end{equation}

The existence of solutions of the equations 
$\mathcal{D}^\dag u(\boldsymbol{r})=0$ 
and 
$\mathcal{D}v(\boldsymbol{r})=0$ 
and the number of solutions of each
kind are determined by the topological properties of the background
fields, $\boldsymbol{A}^{\ }_{5}(\boldsymbol{r})$ and $\varphi(\boldsymbol{r})$.  An index theorem
implies
\begin{equation} 
\label{eq:index} 
\mathrm{Index}(H)= 
\dim \ker \mathcal{D} 
- 
\dim \ker \mathcal{D}^\dag
= 
n
\end{equation} 
where $\ker$ denotes kernel and $n$ is the
vorticity defined in Eq.~(\ref{eq:scalar-field-profile}).  
The implication of this index theorem was seen explicitly in
Ref.~\onlinecite{Jackiw:1981} where solutions of 
$\mathcal{D}^\dag u(\boldsymbol{r})=0$ 
and 
$\mathcal{D}v(\boldsymbol{r})=0$ 
were constructed for the case
of the highly symmetric profile of the vector and Higgs fields
given in Eqs.~(\ref{eq:gauge-field-profile}) and
(\ref{eq:scalar-field-profile}).  It was found that, for a given
vorticity, either one or the other of these equations has solutions,
not both. Which equation had solutions depended on the sign of $n$.
It was argued that the number of solutions is given by $|n|$ and,
when $n=\pm1$, the solutions were found explicitly.

The proof of the index theorem (\ref{eq:index}) was given in
Ref.~\onlinecite{Weinberg:1981eu}. The index theorem counts the difference
 indicated in (\ref{eq:index}). It proves that this is so,
 independent
of the details of the profile of the vector and Higgs fields but
with the assumption that, whatever they are, they are obtained by
smooth deformations of the symmetric configurations in
Eqs.~(\ref{eq:scalar-field-profile}) and (\ref{eq:gauge-field-profile}). 
Ref.~\onlinecite{Weinberg:1981eu} 
also presented a proof of a vanishing theorem, that \textit{either}
$\dim\ker \mathcal{D}=0$ 
\textit{or} 
$\dim\ker \mathcal{D}^\dag=0$.
Combined with the index theorem, it implies that
\begin{subequations}
\begin{eqnarray}
n>0&:&\qquad\dim\ker \mathcal{D}^\dag = 0,~~~
~~\dim\ker \mathcal{D}= n, 
\\
n<0&:&\qquad\dim\ker \mathcal{D}^\dag = |n|,~
~~\dim\ker \mathcal{D} =0 .
\end{eqnarray}
\end{subequations}

A computation of the spectral asymmetry of the Hamiltonian in a
spirit similar to the one that will be given in the remainder of
this Section was originally presented in Ref.~\onlinecite{Niemi:1984gm}.
Equation (6.29) of that paper contains a result for the spectral
asymmetry from which the index can be deduced by taking the
parameter $\kappa$ (our $\mu$) to zero and which agrees with
Eq.~(\ref{eq:index}) above. The general formula for the spectral
asymmetry in their equation (6.29) also agrees with what we shall
find in the following.

\subsection{Non-zero mode spectrum}

Now, we shall look for eigenspinors of the Dirac Hamiltonian which do not have eigenvalues $E=\pm \mu$.

{}From Eq.~(\ref{eq:v}) we can solve for the lower components of the
spinor in terms of the upper components
\begin{equation}
v^{\ }_E(\boldsymbol{r})= 
\frac{1}{E-\mu }\mathcal{D}^\dag u^{\ }_E(\boldsymbol{r}).
\end{equation}
Then, using (\ref{eq:u}) we see that the upper components must obey
the Schr\"odinger equation 
$
\mathcal{D}\mathcal{D}^\dag u^{\ }_E(\boldsymbol{r})=
\left(E^2-\mu ^2\right)u^{\ }_E(\boldsymbol{r})
$.  
To find solutions, we begin
with the eigenvalue problem
\begin{equation}\label{eq:Schrequation}
\mathcal{D}\mathcal{D}^\dag u^{\ }_\lambda(\boldsymbol{r})= 
\lambda u^{\ }_\lambda(\boldsymbol{r}),
\qquad 
\lambda\geq 0.
\end{equation}
We assume that we can find a complete orthornormal set of solutions
of this equation,
\begin{equation}\label{eq:orthonormal}
\int d^2r ~u^{\dag}_\lambda(\boldsymbol{r}) u^{\ }_{\lambda'}(\boldsymbol{r})=
\delta^{\ }_{\lambda\lambda'},
\qquad
\sum_\lambda u^{\ }_\lambda(\boldsymbol{r})u^{\dag}_\lambda(\boldsymbol{r} ')=
\delta(\boldsymbol{r}-\boldsymbol{r} ')\,\openone.
\end{equation}
Generally, the spectrum will contain both bound and continuum
states. For continuous spectra, the right-hand side of the first
equation above should be replaced with a Dirac delta function and
the summation on the left of the second equation should be replaced
by an integral. We shall assume that these replacements, where
needed, are understood in Eq.~(\ref{eq:orthonormal}). We can use the
two-component spinor $u^{\ }_\lambda(\boldsymbol{r})$ to construct a normalized
four-component spinor which solves the stationary Dirac equation.  
For each eigenvalue $\lambda$,
\begin{subequations}
\begin{eqnarray}
\label{eq:spinor1}
\Psi^{\ }_{E}(\boldsymbol{r}) = 
\left( 
\frac{
\sqrt{\lambda+\mu^2}-\mu 
     }
     {
2\sqrt{\lambda+\mu^2}
     }
\right)^{\frac{1}{2}}
\begin{pmatrix} 
u^{\ }_\lambda(\boldsymbol{r}) 
\\ 
\frac{\mathcal{D}^\dag}{ \sqrt{\lambda+\mu^2}-\mu}u^{\ }_\lambda(\boldsymbol{r})
\\
\end{pmatrix},
\qquad
E= \sqrt{\lambda+\mu^2},
\\
\label{eq:spinor2} 
\Psi^{\ }_{E}(\boldsymbol{r}) = 
\left( 
\frac{
\sqrt{\lambda+\mu^2}+\mu }
     {
2\sqrt{\lambda+\mu^2}
     }\right)^{\frac{1}{2}}
\begin{pmatrix} 
u^{\ }_\lambda(\boldsymbol{r}) 
\\ 
\frac{-\mathcal{D}^\dag}{ \sqrt{\lambda+\mu^2}+\mu}
u^{\ }_\lambda(\boldsymbol{r})
\end{pmatrix},
\qquad
E= -\sqrt{\lambda+\mu^2}.
\end{eqnarray}
\end{subequations}
For every $u^{\ }_\lambda$ which is a solution of the Schr\"odinger
equation (\ref{eq:Schrequation}) with  positive eigenvalue
$\lambda>0$, we obtain two solutions of the Dirac equation, one with
positive energy, $E=\sqrt{\lambda+\mu^2}$ and one with negative
energy $E=-\sqrt{\lambda+\mu^2}$. Unlike the zero modes that we
discussed in the previous Subsection, where there was either a
positive or a negative energy solution, here, the positive and
negative energy solutions of the Dirac equation are paired: for each
positive energy solution there is a negative energy solution and
vice-versa. This implies that, if there are bound states other than
the zero modes, they must occur in positive and negative energy
pairs.  Thus, bound states, other than the zero modes, will not
contribute to the spectral asymmetry.  We will see this explicitly
in the following. However, for states in the continuum spectrum, the
pairing tells us only that the spectrum occurs symmetrically about
zero: for example, there is continuum spectrum in the symmetrically
placed regions $E>\sqrt{m^2+\mu^2}$ and $E<-\sqrt{m^2+\mu^2}$. It
does not tell us about the density of states in these regions, which
can still be asymmetric.

\subsection{Charge density}

Let us examine the charge density of the ground state of the system
that we are considering. The charge density is given in
Eq.~(\ref{eq: final result sec QM})
\begin{equation}\label{eq:charge1}
\rho(\boldsymbol{r})= 
\frac{1}{2}u^{\dag}_{-\mu}(\boldsymbol{r}) u^{\ }_{-\mu}(\boldsymbol{r})
-
\frac{1}{2}v^{\dag}_{\mu}(\boldsymbol{r})v^{\ }_{\mu}(\boldsymbol{r})
-
\frac{1}{2}\sum_{E\neq \pm\mu} 
\mathrm{sign}(E)\Psi^{\dag}_{E}(\boldsymbol{r})\Psi^{\ }_E(\boldsymbol{r}).
\end{equation}
Here, we have included both types of zero modes.  Depending on the
sign of the vorticity, only one of them will be non-zero and will
have mutliplicity given by the magnitude of the vorticity.  A sum
over these degenerate wave functions is implied in the first two
terms on the right-hand side of (\ref{eq:charge1}).  We have also
assumed that $\mu$ is positive, so that $v^{\ }_\mu$ is a positive energy
state and $u^{\ }_{-\mu}$ is a negative energy state.  We shall restore
the possibility that $\mu$ could have a negative sign later, where
it will simply lead to a flip in sign from the contribution of the
zero modes. 

Now, using Eqs.~(\ref{eq:spinor1}) and (\ref{eq:spinor2}), 
we find that
the third term in the right-hand side of (\ref{eq:charge1}) is
\begin{eqnarray}\label{eq:ch0}
\rho(\boldsymbol{r})&=& 
\frac{1}{2}u^{\dag}_{-\mu}(\boldsymbol{r}) u^{\ }_{-\mu}(\boldsymbol{r})
-
\frac{1}{2}v_{\mu}^\dag(\boldsymbol{r})v^{\ }_{\mu}(\boldsymbol{r})
\nonumber\\
&&
+
\sum_{\lambda>0}\frac{\mu}{2\sqrt{\lambda+\mu^2}} 
\left(
u_\lambda^\dag(\boldsymbol{r})
u^{\ }_\lambda(\boldsymbol{r}) 
- 
\frac{1}{\lambda}
\left(\mathcal{D}^\dag u^{\ }_\lambda(\boldsymbol{r})\right)^\dag 
\mathcal{D}^\dag u^{\ }_\lambda(\boldsymbol{r})
\right).
\end{eqnarray}
Using the fact that $u^{\ }_\lambda$ satisfies the Schr\"odinger equation
(\ref{eq:Schrequation}) leads to
\begin{eqnarray}
\label{charge}
\rho(\boldsymbol{r})&=& 
\frac{1}{2}u^{\dag}_{-\mu}(\boldsymbol{r}) u^{\ }_{-\mu}(\boldsymbol{r})
-
\frac{1}{2}v_{\mu}^\dag(\boldsymbol{r})v^{\ }_{\mu}(\boldsymbol{r})
\nonumber\\
&&
+
\sum_{\lambda>0} \frac{\mu }{2\lambda\sqrt{\lambda+\mu^2}} 
\left(
u_\lambda^\dag(\boldsymbol{r})\mathcal{D}\mathcal{D}^\dag u^{\ }_\lambda(\boldsymbol{r})
- \left( \mathcal{D}^\dag u^{\ }_\lambda(\boldsymbol{r})\right)^\dag {\cal
D}^\dag u^{\ }_\lambda(\boldsymbol{r})
\right).
\end{eqnarray}
The last terms in this expression are a total derivative
\begin{eqnarray} 
\label{eq:finalexpressionforchargedensity}
\rho(\boldsymbol{r})&=& 
\frac{1}{2}u^{\dag}_{-\mu}(\boldsymbol{r})
u^{\ }_{-\mu}(\boldsymbol{r})
-
\frac{1}{2}v_{\mu}^\dag(\boldsymbol{r})v^{\ }_{\mu}(\boldsymbol{r}) 
\nonumber\\
&&
+
\boldsymbol{\partial} \cdot  \sum_{\lambda>0} 
\frac{\mu}{2\lambda\sqrt{\lambda+\mu^2}}
\left( 
u_\lambda^\dag(\boldsymbol{r})
~\mathrm{i}\boldsymbol{\sigma}~
\mathcal{D}^\dag u^{\ }_\lambda(\boldsymbol{r})
\right).
\end{eqnarray}
The total charge is a volume integral of the charge density. If we
volume integrate the last term in the equation above and use Gauss'
theorem, it will be expressed as a line integral on the circle at
infinity of the quantity which is to the right of the derivative
operator.  Thus we see that the charge will depend on the asymptotic
form of the wave-functions.   We observe that, consistent with our
discussion after Eqs.~(\ref{eq:spinor1}) and (\ref{eq:spinor2}),
since the  wave-functions of bound states fall off exponentially at
large distances, bound states will not contribute to the charge.
Only continuum states are important.   Further, studying the
asymptotics of the continuum states will allow us to compute the
total charge. What will make the task easy is the fact that the
volume integral of the part of the last term in
(\ref{eq:finalexpressionforchargedensity}) will pick up
contributions  which go like $1/r$.

Before we do that, we re-organize the expression for the charge
density. We use the identity
\begin{equation}
\frac{\mu}{2\sqrt{\lambda+\mu^2}}=\int_{-\infty}^\infty \frac{d\omega}{2\pi}
\frac{\mu}{\lambda+\mu^2+\omega^2}
\end{equation}
 and the Schr\"odinger equation (\ref{eq:Schrequation}) 
to re-write Eq.~(\ref{charge}) as
\begin{equation}
\label{eq:ch} 
\begin{split}
\rho(\boldsymbol{r})=&\,
\frac{1}{2}u^{\dag}_{-\mu}(\boldsymbol{r})u^{\ }_{-\mu}(\boldsymbol{r})
-
\frac{1}{2}v_{\mu}^\dag(\boldsymbol{r})v^{\ }_{\mu}(\boldsymbol{r}) 
\\
&\,
+
\boldsymbol{\partial} \cdot \int_{-\infty}^\infty\frac{d\omega}{2\pi}
\left(
\sum_{\lambda>0}u_\lambda^\dag(\boldsymbol{r})
\frac{1}{\mathcal{D}\mathcal{D}^\dag}\frac{\mu}{\mathcal{D}\mathcal{D}^\dag
+
\mu^2+\omega^2}
~\mathrm{i}\boldsymbol{\sigma}~
\mathcal{D}^\dag u^{\ }_\lambda(\boldsymbol{r})
\right),
\end{split}
\end{equation}
or, 
as the basis-independent expression
\begin{equation}
\begin{split}
\label{eq:ch1} 
\rho(\boldsymbol{r})=&\,
\frac{1}{2}u^{\dag}_{-\mu}(\boldsymbol{r})u^{\ }_{-\mu}(\boldsymbol{r})
-
\frac{1}{2}v_{\mu}^\dag(\boldsymbol{r})v^{\ }_{\mu}(\boldsymbol{r}) 
\\
&\,
+\boldsymbol{\partial} \cdot \int_{-\infty}^\infty\frac{d\omega}{2\pi}
~\mathrm{tr}~ 
\left\langle\boldsymbol{r}\left|
\frac{P}{\mathcal{D}\mathcal{D}^\dag}\,
\frac{\mu}{\mathcal{D}\mathcal{D}^\dag +\mu^2+\omega^2}
~\mathrm{i}\boldsymbol{\sigma}~
\mathcal{D}^\dag 
\right|\boldsymbol{r}\right\rangle,
\end{split}
\end{equation}
where ``tr'' denotes a trace over Dirac matrices and $P$ is a
projection operator onto states orthogonal to the zero mode
wave-functions. (We shall make use of the expression only where
$|\boldsymbol{r}|\to \infty$ and the zero-mode wave-functions have vanishing
contribution. For this reason, explicit use of this projection will
never be needed.)

\textbf{Case} $\boldsymbol{A}^{\ }_{5}=0$:
Let us first consider the case where the axial vector gauge field is absent.  In
the asymptotic region, the Higgs field is
\begin{subequations}
\begin{eqnarray} 
\label{scalarfieldasymptotic}
&&
\varphi(\boldsymbol{r})\equiv
\varphi^{\ }_{1}(\boldsymbol{r})+\mathrm{i}\varphi^{\ }_{2}(\boldsymbol{r})=
{\phi}\; e^{\mathrm{i}n\theta}+\mathcal{O}(r^{-2}),
\\
&&
\partial^{\ }_{i}\varphi(\boldsymbol{r})=
-\mathrm{i}n\epsilon^{\ }_{ij}\frac{r^j}{r^2}~{\phi}\; e^{\mathrm{i}n\theta}+\mathcal{O}(r^{-3}).
\end{eqnarray} 
\end{subequations}
With
\begin{eqnarray}\label{eq:DDdagger}\mathcal{D}\mathcal{D}^\dag =-
\partial_1^2  -
\partial_2^2+|\varphi|^2
+
\boldsymbol{\sigma}\cdot\boldsymbol{\partial} 
(\varphi^{\ }_{1}+\mathrm{i}\sigma^{\ }_{3}\varphi^{\ }_{2}),
\end{eqnarray}
we see that the derivatives of the Higgs field provide a
long-ranged potential $\sim 1/r$ in the Schr\"odinger equation
(\ref{eq:Schrequation}).

It is easy to find the asymptotic behavior of the propagators in
Eq.~(\ref{eq:ch1}) by perturbative expansion in the deviation of the
operator in Eq.~(\ref{eq:DDdagger}) from the free operator  ${\cal
D}\mathcal{D}^\dag|^{\ }_{\varphi=m}=-
\partial_1^2  -
\partial_2^2+m^2 $. 
For example,
\begin{equation}
\begin{split}
&
\left\langle\boldsymbol{r}\left|
\frac{1}{\mathcal{D}\mathcal{D}^\dag+\mu^2+\omega^2}
\right|\boldsymbol{r}'\right\rangle=
\\ 
&
\qquad
\int \frac{d^2p}{(2\pi)^2}e^{\mathrm{i}\boldsymbol{p}\cdot(\boldsymbol{r}-\boldsymbol{r}')}  
\frac{1} {p^2+{\phi}^2+\mu^2+\omega^2}
\\ 
&
\qquad
-
\boldsymbol{\sigma}
\cdot
\boldsymbol{\partial} 
\Big(
\varphi^{\ }_{1}(\boldsymbol{r})
+
\mathrm{i}\sigma^{\ }_{3}\varphi^{\ }_{2}(\boldsymbol{r})
\Big)
\int \frac{d^2p}{(2\pi)^2}
e^{\mathrm{i}\boldsymbol{p}\cdot(\boldsymbol{r}-\boldsymbol{r}')} 
\frac{1}{[p^2+{\phi}^2+\mu^2+\omega^2]^2 }+\cdots.
\label{eq:propagatorasymptotic}
\end{split}
\end{equation}
The right-hand side in this equation has support in the region where
${\phi} |\boldsymbol{r} - \boldsymbol{r}'|<1$ 
as it falls off exponentially with the distance
$|\boldsymbol{r} - \boldsymbol{r}'|$
when this distance is greater than $1/{\phi}$. 
The second-term on the right-hand side of
(\ref{eq:propagatorasymptotic}) goes like $1/r$ and the corrections
to it, represented by $\cdots$, fall off faster and will not be
needed.

Using the asymptotic expression (\ref{eq:propagatorasymptotic}) in
Eq.~(\ref{eq:ch1}), we obtain
\begin{equation}  
\begin{split}
\rho(\boldsymbol{r})=&\,
\frac{1}{2}u^{\dag}_{-\mu}(\boldsymbol{r})
u^{\ }_{-\mu}(\boldsymbol{r})
-
\frac{1}{2}v_{\mu}^\dag(\boldsymbol{r})v^{\ }_{\mu}(\boldsymbol{r}) 
\\
&\,
-
\boldsymbol{\partial}
\wedge
\left(
\frac{\mu}{8\pi {\phi}^2\sqrt{m^2+\mu^2}}
\left(
\varphi^*\mathrm{i}\boldsymbol{\partial}\varphi 
-
\mathrm{i}\boldsymbol{\partial}\varphi^*\varphi
\right) 
+
\cdots
\right).
\end{split}
\end{equation}
Upon integrating this expression, we obtain
\begin{equation}
Q=-\frac{1}{2}~\mathrm{Index}(H)~ 
-\frac{\mu}{2\sqrt{{\phi}^2+\mu^2}}\frac{1}{4\pi m^2}
\oint d\boldsymbol{l}\cdot
\left(
\varphi^*\mathrm{i}\boldsymbol{\partial}\varphi 
- 
\mathrm{i}\boldsymbol{\partial}\varphi^*\varphi
\right)
\end{equation}
where, in the first term in the right-hand side,  we have remembered
that the number of zero modes is determined by the index and the
line integral in the second term is taken on the circle at infinite
radius. Using the index theorem~(\ref{eq:index}) and the asymptotic
expression for $\varphi$ in Eq.~(\ref{scalarfieldasymptotic}), we
obtain
\begin{equation}
Q=
\left(-\frac{1}{2}~\mathrm{sign}(\mu)~ 
+
\frac{\mu}{2\sqrt{{\phi}^2+\mu^2}}\right)n.
\label{eq:ch3} 
\end{equation} 
We have restored the possibility that
$\mu$ could be negative in the first term by recalling that the sign
of the energy of the zero modes is determined by $\mu$.

\textbf{Case} $\boldsymbol{A}^{\ }_{5}\neq0$:
The second case is when there is also 
an axial vector gauge field with asymptotic form
\begin{eqnarray}\label{vectorfieldasymptotic}
A^{i}_{5}(\boldsymbol{r})=
-n\epsilon^{ij}\frac{r^j}{2r^2}
+
\mathcal{O}(r^{-2})
\end{eqnarray}
so that the covariant derivative of the Higgs field falls of at infinity
\begin{equation}
\left(\boldsymbol{\partial}+2\mathrm{i}\boldsymbol{A}^{\ }_{5}\right)
\varphi(\boldsymbol{r}) 
= \mathcal{O}(r^{-2}).
\end{equation} 
Here, we have assumed a power-law
fall-off that is sufficiently fast for our purposes. 
In fact, for a classical field theory with a vortex solution, 
the covariant derivative falls off exponentially with distance from the vortex.
Then,
\begin{equation}
\left\langle\boldsymbol{r}\left|
\frac{1}{\mathcal{D}\mathcal{D}^\dag+\mu^2+\omega^2}
\right|\boldsymbol{r}'\right\rangle=
e^{
-
\mathrm{i}\sigma^{\ }_{3}\int\limits_{r}^{r'} 
d\boldsymbol{\ell}\cdot\boldsymbol{A}^{\ }_{5}
  }
\int
\frac{d^2p}{(2\pi)^2}
e^{
\mathrm{i}\boldsymbol{p}\cdot(\boldsymbol{r}-\boldsymbol{r}')
  }  
\frac{1}
     {p^2+{\phi}^2+\mu^2+\omega^2}
+
\cdots.
\label{eq:asymptoticforAneq0}
\end{equation}
Corrections represented by $\cdots$ fall off at least as fast as $1/r^2$
as $r\to\infty$. The line integral in the phase factor is to be
taken along a straight line between $\boldsymbol{r}$ and $\boldsymbol{r}'$. 
(Since the axial magnetic field also goes to zero at least as fast as $r^{-2}$ 
as $r\to\infty$, the path is not important for our purposes.)

The trivial asymptotic form of (\ref{eq:asymptoticforAneq0}) means
that the background fields do not contribute to the relevant
asymptotic of the last term in (\ref{eq:ch1}) and the volume
integral of that term vanishes.  It therefore does not contribute to
the total charge. We find that the charge in this case is entirely
determined by the zero modes,
\begin{equation}
Q=-
\frac{1}{2}~\mathrm{Index}(H)~\mathrm{sign}(\mu)=-
\frac{n}{2}~\mathrm{sign}(\mu).
\end{equation}

This is dramatically different from the result for the case without
an axial vector gauge field quoted in Eq.~(\ref{eq:ch3}). 
As we have seen, the
difference can be attributed to the asymptotics of the background
field configurations.  Another way to understand it is to realize
that, when the Higgs field approaches its asymptotic form its
covariant derivatives as well as the axial magnetic field fall off
quickly enough at $r\to\infty$, so that stereographic projection can be used
to map the problem of solving the Dirac equation on the plane to the
problem of solving it on the sphere (where the vector field is a
connection on a Wu-Yang monopole bundle). Then, the entire spectrum
is discrete and, by the arguments following Eqs.~(\ref{eq:spinor1})
and (\ref{eq:spinor2}) we can see that all non-zero-mode solutions
of the Schr\"odinger equation with eigenvalue $\lambda$ result in
pairs of solutions of the Dirac equation: one positive
$E=\sqrt{\lambda+\mu^2}$ and one negative energy
$E=-\sqrt{\lambda+\mu^2}$ state. For this reason, only the zero
modes can contribute to the spectral asymmetry and the contribution
must be proportional to the index.

As is well known, the axial gauge field is needed to render the vortex energy finite; it screens the infinite energy coming from the scalar field. 
Evidently, it also screens the irrational charge which arises
from the staggered chemical potential. Some further insight into this phenomenon is given below.

\section{
Field theory analysis
        }
\label{sec: Field theory analysis}

An alternative method for finding the charge induced by the vortex
background makes use of a field theoretic evaluation of the
expectation value of the current in the ``vacuum'' state for the Dirac
field operators in the vortex background,
\begin{equation}
J^\nu(x) =
\langle \bar{\psi}(x) \gamma^\nu \psi(x) \rangle = 
- 
\mathrm{Tr}\left[\gamma^\nu S(x,x)\right],
\end{equation}
where $S(x,y)$ is the Dirac field propagator  
for the Lagrange density~(\ref{eq:Lagrangian}).

We consider first the theory without the axial gauge field and present
Eq.~(\ref{eq:Lagrangian}) as
\begin{subequations}
\begin{equation}
\label{eq:Lagrangian-no-gauge}
\mathcal{L}^{\ }_{0}=
\bar{\psi} 
\big(\mathrm{i} \gamma^\nu \partial^{\ }_\nu - \Phi \big)\psi
\end{equation}
where
\begin{equation}
\Phi =
\varphi^{\ }_{1} 
- 
\mathrm{i}
\gamma^5 
\varphi^{\ }_{2} 
+ 
\gamma^3 \mu\equiv
\varphi^{\ }_{1} 
- 
\mathrm{i}
\gamma^5 
\varphi^{\ }_{2} 
+ 
\gamma^3 
\varphi^{\ }_{3}.
\label{eq:Phi-mass}
\end{equation}
\end{subequations}
Evidently, we need to invert
\begin{equation}
\label{eq:propagator-no-gauge}
S^{-1}(x,y) =
-\mathrm{i} 
\left( \mathrm{i} \gamma^\nu \partial^{\ }_\nu -\Phi \right) \delta(x-y)\; .
\end{equation}
This can be done perturbatively in a gradient expansion for $\Phi$.
We set 
\begin{equation}
\Phi(x)= 
M+\delta\Phi(x), 
\qquad
M=\Phi(0),
\qquad
\delta\Phi(x) =x^\nu \partial^{\ }_\nu \Phi(0). 
\end{equation}
In a graphical representation, a thick
line denotes $S(x,y)$ 
while a thin line represents the free propagator.
\begin{equation}
S^{\ }_0(x,y)= 
\int\frac{d^3 p}{(2\pi)^3} 
\frac{\mathrm{i}}{\slashed{p}-M}\,
e^{\mathrm{i}p(x-y)},
\qquad
\slashed{p}\equiv\gamma^{\nu}p^{\ }_{\nu}.
\end{equation}
Hence we have
\begin{equation}
\includegraphics[width=0.60\linewidth]{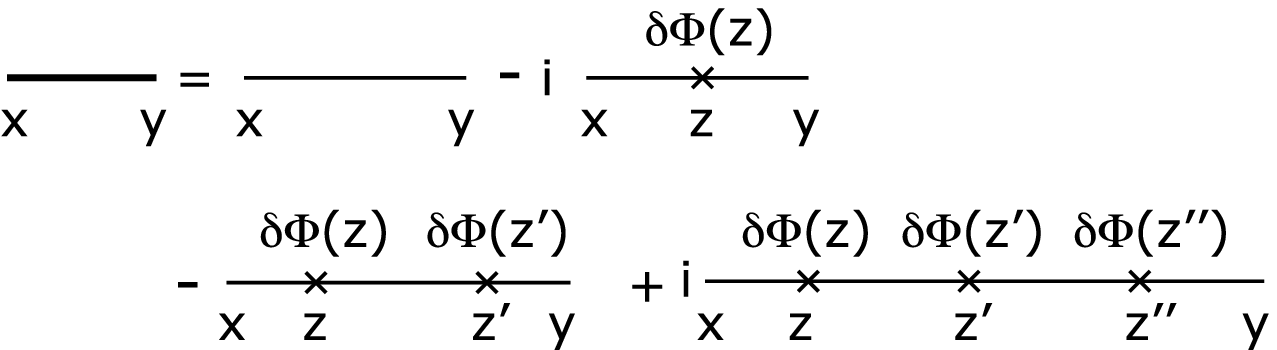}
\label{eq:diagram-S}
\end{equation}
with integration over the $z$-variables understood. 
The result of the calculation is
\begin{subequations}
\begin{equation}
J^{\nu}=
\frac{-1}{8\pi m^3}
\epsilon^{\nu\alpha\beta}\;
\epsilon^{abc}\; 
\varphi^{\ }_a\,
\partial^{\ }_\alpha\varphi^{\ }_b\,
\partial^{\ }_\beta \varphi^{\ }_c=
\frac{-1}{8\pi}
\epsilon^{\nu\alpha\beta}\;
\epsilon^{abc}\; 
n^{\ }_a\,
\partial^{\ }_\alpha n^{\ }_b\,
\partial^{\ }_\beta n^{\ }_c
\label{eq:current-no-gauge a}
\end{equation}
where 
\begin{equation}
m^{2}=
\sum_{a=1}^{3}
\varphi^{2}_{a}=
\phi^{2}+\mu^{2},
\qquad
n^{\ }_{a}=
\frac{\varphi^{\ }_{a}}{m},
\qquad a=1,2,3.
\label{eq:current-no-gauge b}
\end{equation}
\end{subequations}
The second equality in 
Eq.~(\ref{eq:current-no-gauge a}) 
shows that $J^{\nu}$ is manifestly divergence-free. [The overall sign is set by the requirement that the isolated gap state is filled.]

To evaluate the induced charge, we observe that the charge density for
static fields is
\begin{equation}
\rho= \frac{-1}{ 8 \pi m^3 } 
\;\epsilon^{ij} 
\;\epsilon^{abc} 
\;\varphi^{\ }_a 
\;\partial^{\ }_i \varphi^{\ }_b 
\;\partial^{\ }_j \varphi^{\ }_c.
\end{equation}
With our profile Eq.~(\ref{eq:scalar-field-profile}), this becomes
\begin{equation}
\rho(r) = 
 \frac{n}{4 \pi r} \frac{d}{dr} \frac{ \mu(r)}{ m(r)}
\end{equation}
whose spatial integral yields
\begin{equation}
\label{eq:irrational-Q}
Q=
\int d^2r\, \rho(r) = 
 \left.\frac{n}{2} \frac{\mu(r)}{m(r)}\right|_0^\infty 
=
(-\frac{1}{2}\; \mathrm{sign}\big(\mu(0)\big)
+\frac{1}{2}\; \frac{\mu(\infty)}{m(\infty)})n,
\end{equation}
since the amplitude $\phi$ vanishes at the origin.
This charge can be an irrational quantity, reducing to $\pm{n}/{2}$
as the staggered chemical potential tends to $\pm 0$.

It is noteworthy that the induced current~(\ref{eq:current-no-gauge a}) 
exhibits an $SO(3)$
algebraic structure, even though neither the Lagrange density in
Eq.~(\ref{eq:Lagrangian-no-gauge}) nor the propagator in
Eq.~(\ref{eq:propagator-no-gauge}) put such structure into evidence. We
shall explain below how this comes about.

Another interesting point is that the current can take a simpler form
after fields are redefined. First, we rewrite
Eq.~(\ref{eq:current-no-gauge a}) 
in terms of $\varphi$ and $\varphi^{*}$
\begin{equation}
\label{eq:current-no-gauge-1}
J^\nu= 
\frac{ \mathrm{i} }{8\pi m^3}\,
\epsilon^{\nu\alpha\beta}\;
\left[ 
\mu \,
\left(\partial^{\ }_\alpha \varphi\right)^{*}
\,
\left(\partial^{\ }_\beta \varphi\right) 
- 
\partial^{\ }_\alpha \mu\,(\varphi^* \partial^{\ }_\beta \varphi 
- \varphi \partial^{\ }_\beta \varphi^*)  \right].
\end{equation}
Next, we define
\begin{equation}
\label{eq:definition-chi}
\varphi = 2 m \chi \sqrt{1- |\chi |^2}, \qquad \mu= m \left(1- 2 |\chi |^2\right),
\end{equation}
thereby expressing the current~(\ref{eq:current-no-gauge-1}) as
\begin{equation}
J^\nu= 
\frac{ \mathrm{i} }{2\pi } 
\epsilon^{\nu\alpha\beta}\,
\partial^{\ }_\alpha \chi^* 
\,\partial^{\ }_\beta \chi=
\frac{\mathrm{i}}{4\pi }  
\epsilon^{\nu\alpha\beta} 
\partial^{\ }_{\alpha} 
\left( 
\chi^* 
\partial^{\ }_\beta 
\chi
- 
\chi 
\partial^{\ }_\beta 
\chi^*
\right).
\end{equation}
This shows that $J^\nu$ is a total divergence and is manifestly conserved.

Next we write the current when the axial gauge potential is present. 
Rather than calculating from first principles, we appeal to local
axial gauge invariance, and promote all the derivatives in 
Eq.~(\ref{eq:current-no-gauge-1}) 
to covariant derivatives
\begin{equation}
D^{\ }_\nu \equiv 
\partial^{\ }_\nu 
+
2\mathrm{i} A^{\ }_{5\nu}.
\end{equation} 
However, the resulting expression is not
conserved, but it can be made conserved by adding an axial gauge invariant
term, which is linear in the axial gauge field. In this way we arrive at
\begin{equation}
\label{eq:current-gauge}
J^\nu 
=  
\frac{ \mathrm{i} }{ 8 \pi m^3} 
\;\epsilon^{\nu \alpha \beta} 
\;\left\{ 
\mu\, (D^{\ }_\alpha \varphi)^* (D^{\ }_\beta \varphi) 
- \partial^{\ }_\alpha \mu \,[ \varphi^* (D^{\ }_\beta \varphi) 
- \varphi (D^{\ }_\beta \varphi^*) ]  
\right\}
+\frac{1}{2\pi} \frac{\mu}{m}\, {F}^\nu_5,
\end{equation}
where ${F}^\nu_5$ is the axial dual field strength 
\begin{equation}
{F}^\nu_5 \equiv 
\frac{1}{2} \,
\epsilon^{\nu \alpha \beta} \,F^{\ }_{5\alpha \beta} 
= 
\epsilon^{\nu \alpha \beta} \partial^{\ }_\alpha A^{\ }_{5\beta}.  
\end{equation}
As a check, the coefficient of the last term in
Eq.~(\ref{eq:current-gauge}) can be computed from the relevant
graph. When the axial gauge potential contribution to
Eq.~(\ref{eq:current-gauge}) is separated, 
Eq.~(\ref{eq:current-gauge}) equals
\begin{equation}
J^\nu 
=  \frac{\mathrm{i} }{ 8 \pi m^3} 
\;\epsilon^{\nu \alpha \beta} 
\;\left\{ \mu \, (\partial^{\ }_\alpha \varphi)^* (\partial^{\ }_\beta \varphi) 
- \partial^{\ }_\alpha \mu \,[ \varphi^* (\partial^{\ }_\beta \varphi) 
- \varphi (\partial^{\ }_\beta \varphi^*) ]  
\right\}
+ \frac{ \epsilon^{\nu \alpha \beta} }{ 2\pi } 
\partial^{\ }_\alpha \left( \frac{\mu}{m} A^{\ }_{5\beta}\right) .
\end{equation}
Therefore, the axial gauge potential's contribution to the charge density is 
\begin{equation}
\Delta \rho =  
- \frac{\epsilon^{ij}}{2 \pi} \,\partial^{\ }_i \left(\frac{\mu}{m} A^j_5\right),
\end{equation}
which, for $A^{\nu}_5$ 
as in Eq.~(\ref{eq:gauge-field-profile}), equals 
\begin{equation}
\Delta \rho(r) = - \frac{n}{2 \pi} \;
\frac{1}{r} \frac{d}{d r} \left( \frac{\mu(r)}{m(r)} a^{\ }_{5}(r) \right),
\end{equation}
and its contribution to the total charge is
\begin{equation}
\Delta Q 
=
\int d^2r \;\Delta \rho(r)
= 
- \frac{n}{2}\; \frac{\mu(\infty)}{m(\infty)}
\end{equation}
since $a^{\ }_{5}(\infty)=1/2$ while $a^{\ }_{5}(0)=0$. 
This cancels the continuous dependence on $\mu(\infty)$ found
in Eq.~(\ref{eq:irrational-Q}), leaving the same rational result
obtained in the absence of the staggered chemical potential.

Note that with variables defined as in Eq.~(\ref{eq:definition-chi}), 
the current in the presence of the axial gauge field reads
\begin{eqnarray}
J^\nu&=& \frac{\mathrm{i}}{2\pi }\,
\epsilon^{\nu\alpha\beta}
(D^{\ }_\alpha \chi)^* 
(D^{\ }_\beta  \chi) 
+
\frac{1}{2\pi}
(1-2|\chi|^2)\,
F^\nu_5
\nonumber
\\
&=& \frac{ \mathrm{i} }{4\pi }
\epsilon^{\nu\alpha\beta} 
\partial^{\ }_\alpha
\left[
\chi^{*}
D^{\ }_{\beta}
\chi
-
\chi
D^{\ }_{\beta}
\chi^{*}
-
2\mathrm{i}A^{\ }_{5\beta}
\right],
\end{eqnarray}
or when the gauge field is separated
\begin{eqnarray}
J^\nu&=& \frac{ \mathrm{i} }{2\pi }
\epsilon^{\nu\alpha\beta} 
(\partial^{\ }_\alpha \chi)^* 
(\partial^{\ }_\beta \chi) 
+
\frac{1}{2 \pi}
\epsilon^{\nu\alpha\beta}
\partial^{\ }_\alpha
\Big[ 
(1-2|\chi|^2)
A^{\ }_{5\beta} 
\Big]\nonumber
\\
&=& \frac{ \mathrm{i} }{4\pi  }
\epsilon^{\nu\alpha\beta} 
\partial^{\ }_\alpha
\left[
\chi^{*}
\partial^{\ }_{\beta}
\chi
-
\chi
\partial^{\ }_{\beta}
\chi^{*}
-
2\mathrm{i}
\left(
1
-
2|\chi|^{2}
\right)
A^{\ }_{5\beta}
\right].
\end{eqnarray}
Therefore, the total divergence feature 
and the conservation of the current are again explicitly
exhibited also in the presence of the axial gauge field. 
In particular for the charge density, we have
\begin{equation}
\rho 
=  \frac{ \mathrm{i} }{4\pi} \;\epsilon^{ij}
\partial^{\ }_i [ \chi^* (D^{\ }_j \chi) - \chi (D^{\ }_j \chi)^*] 
+ \frac{1}{2 \pi}\;\epsilon^{ij} \partial^{\ }_i A^{\ }_{5j}.
\end{equation}
Upon integration over space, the first term is cast on the circle at
infinity, where the covariant derivatives of $\chi$ vanish. The second
term shows that the induced charge is exactly the vortex flux, equal
to $n/2$ for $\boldsymbol{A}^{\ }_{5}$ 
as in Eq.~(\ref{eq:gauge-field-profile}).

\section{
Induced charge from symmetry arguments
        }
\label{sec: Induced charge from symmetry arguments}

As we observed in Sec.~\ref{sec: Field theory analysis}, 
the form of the induced current%
~(\ref{eq:current-no-gauge a}) 
without an axial gauge field $A^{\nu}_{5}$
exhibits an $SO(3)$ algebraic structure
despite the absence of any such symmetry
in the Lagrange density~(\ref{eq:Lagrangian-no-gauge}).%
~\cite{footnote on SU(2) sym}
In this section we explain why this is so.
Also we obtain expressions for the induced
fractional charge from symmetry arguments.

Our starting point is the Lagrange density%
~(\ref{eq:Lagrangian})
with the mass terms collected into $\Phi$,
as in~(\ref{eq:Phi-mass}),
but written as
\begin{subequations}
\begin{equation}
\mathcal{L}^{\ }_\psi=
\bar\psi
\left[
\gamma_{\ }^{\nu}
\left(
\mathrm{i}\partial^{\ }_{\nu}
+
\gamma^{\ }_{5}A^{\ }_{5\nu}
\right)
-
m
M^{\ }_{a}
n^{\ }_{a}
\right]
\psi
\label{eq: def cal L Psi}
\end{equation}
where
\begin{equation}
M^{\ }_1=\openone,
\qquad
M^{\ }_2=-\mathrm{i}\gamma^5, 
\qquad
M^{\ }_3=\gamma^3. 
\end{equation}
\end{subequations}
The background fields
$mn^{\ }_{a}$,
functions on two-dimensional $(2+1)$ dimensional space-time,
are given by
\begin{equation}
n^{\ }_{1}=\frac{\varphi^{\ }_{1}}{m},
\qquad
n^{\ }_{2}=\frac{\varphi^{\ }_{2}}{m},
\qquad
n^{\ }_{3}=\frac{\mu}{m},
\end{equation}
which, since $m^2=|\varphi|^2 + \mu^2$, 
satisfy the local constraint
\begin{equation}
1
=
n^{2}_{1}
+
n^{2}_{2}
+
n^{2}_{3}
\equiv
\boldsymbol{n}^{2}.
\label{eq:2-sphere}
\end{equation}
Despite the suggestive form in which the 3-dimensional vector
$\boldsymbol{n}$ is written above, the Lagrange density~(\ref{eq: def cal L Psi}) is \textit{not} $SU(2)$ 
symmetric because the $M^{\ }_a$ matrices 
do not satisfy the $su(2)$ algebra. 

However, it is the induced current
~(\ref{eq:current-no-gauge a})
and not the starting Lagrange density
~(\ref{eq: def cal L Psi}) that exhibits the symmetry.
Thus, let us turn our attention to the $U(1)$ charge
current induced by the background field
$m\boldsymbol{n}$,
\begin{equation}
\begin{split}
J^{\nu}(x)=&
\frac{
\int\mathrm{D}[\bar\psi,\psi]\
e^{ \mathrm{i}
\int d^{3}x\,
\mathcal{L}^{\ }_{\psi}
  }
\
\big(
\bar\psi
\gamma^{\nu}
\psi
\big)(x)
     }
     {
\int\mathrm{D}[\bar\psi,\psi]\
e^{ \mathrm{i}
\int d^{3}x\,
\mathcal{L}^{\ }_{\psi}
  }
\
\hphantom{
\left(
\bar\psi
\gamma^{\ }_{\nu}
\psi
\right)(x)
         }
     }.
\end{split}
\label{eq: U(1) current if psi and TRS}
\end{equation}
In Eq.~(\ref{eq: U(1) current if psi and TRS}),
we are free to change integration variables as long
as this transformation leaves the current unchanged. 
This we do through the nonunitary transformation
\begin{equation}
\bar\psi=
\bar\chi
\;\gamma^{\ }_{5}\gamma^{3},
\qquad
\psi=
\chi,
\end{equation}
for some arbitrarily chosen constant unit vector $\boldsymbol{N}$.
Thus, the induced current~(\ref{eq: U(1) current if psi and TRS})
is now given by
\begin{subequations}
\begin{equation}
\begin{split}
J^{\nu}=&\,
\frac{
\int\mathrm{D}[\bar\chi,\chi]\
e^{
\mathrm{i}
\int d^{3}x\,
\mathcal{L}^{\ }_{\chi}
  }
\
\big(
\bar\chi
\Gamma^{\nu}
\chi
\big)
     }
     {
\int\mathrm{D}[\bar\chi,\chi]\
e^{ \mathrm{i}
\int d^{3}x\,
\mathcal{L}^{\ }_{\chi}
  }
\
\hphantom{
\left(
\bar\chi
\Gamma^{\nu}
\chi
\right)
         }
     }
\end{split}
\label{eq: U(1) current if chi and TRS}
\end{equation}
where the transformed Lagrange density reads
\begin{eqnarray}
&&
\mathcal{L}^{\ }_{\chi}=
\bar\chi
\left[
\Gamma_{\ }^{\nu}
\left(
\mathrm{i}\partial^{\ }_{\nu}
+
\gamma^{\ }_{5}A^{\ }_{5\nu}
\right)
-
m\Sigma^{\ }_{a}N^{\ }_{a}
\right]
\chi,
\label{eq: cal Lchi}
\end{eqnarray}
with
$\Gamma^{\mu}= \gamma_5\,\gamma^3\,\gamma^{\mu}$
and
$\Sigma^{\ }_{a}= \gamma_5\,\gamma^3\,M^{\ }_{a}$ 
satisfying
\begin{equation}
\{\Gamma^{\ }_{\nu},\Gamma^{\ }_{\nu'}\}=
2\,g^{\ }_{\nu\nu'},
\qquad
[\Sigma^{\ }_{a},\Sigma^{\ }_{b}]=
2\mathrm{i}\epsilon^{\ }_{abc}\; \Sigma^{\ }_{c},
\qquad
[\Gamma^{\ }_{\nu},\Sigma^{\ }_{b}]=0.
\end{equation}
\end{subequations} 
Since $\gamma^{\ }_{5}$ does not commute with all $\Sigma^{\ }_{a}$,
\begin{equation}
\left\{
\Sigma^{\ }_{1},
\gamma^{\ }_{5}
\right\}
=
\left\{
\Sigma^{\ }_{1},
\gamma^{\ }_{5}
\right\}
=
0,
\qquad
\left\{
\Sigma^{\ }_{3},
\gamma^{\ }_{5}
\right\}
=
2\openone,
\qquad
\left[
\Sigma^{\ }_{3},
\gamma^{\ }_{5}
\right]
=
0,
\end{equation}
$\mathcal{L}^{\ }_{\chi}$
is an $SU(2)$ singlet at $A^{\nu}_{5}=0$ only.

\subsection{
Induced charge without axial flux
           }

In the absence of an axial gauge field, 
the Lagrange density~(\ref{eq: cal Lchi})
with $A^{\nu}_{5}=0$ is an $SU(2)$ singlet. 
The induced current and charges must therefore be $SO(3)$
singlets given by
\begin{subequations}
\begin{eqnarray}
&&
J^{\nu}=
C\;\frac{1}{8\pi}\;\epsilon^{\nu\alpha\beta}\,\epsilon^{abc}\;
n^{\ }_a\,\partial^{\ }_\alpha n^{\ }_b\,\partial^{\ }_\beta n^{\ }_c,
\\
&&
Q=\int d^2r\;J^{0}(t,\boldsymbol{r})=
C\;\frac{\Omega}{4\pi},
\label{eq:charge-w-C}
\end{eqnarray}
\end{subequations}
where $\Omega$ is the spherical angle (in units of $4\pi$)
covered by the mapping between the base space
$\boldsymbol{r}\in\mathbb{R}^{2}$ and a closed curve on 
the surface of the 2-sphere~(\ref{eq:2-sphere}),
to lowest order in a gradient expansion.
Thus, we arrived at our previous result, except the factor $C$ must
still be determined.

We can fix the constant $C$ using our results for the fractional
charge derived in the simple case when there is no staggered chemical
potential: the charge is $Q=-1/2$ when the midgap state is empty, and
$Q=1/2$ when the midgap state is filled. There is an ambiguity for the
charge as given in Eq.~(\ref{eq:charge-w-C}) if the bound state is at
exactly zero energy, because then it can be filled or empty, but this
can be lifted by considering the case where $\mu\to 0^{+}$
with $0^{+}$ a positive infinitesimal. In this
case, the bound state solution exists for an antivortex ($n=-1$), and
$E=-\mu$ (see Refs.~\onlinecite{Hou:2007,Jackiw:2007} and
Sec.~\ref{sec: Quantum mechanical analysis}), so that the
level is filled and the charge is thus $Q=1/2$.

Because $\mu\to 0^{+}$ or, equivalently, $n^{\ }_3\to 0^{+}$, the
spherical angle traced by the antivortex in the $n^{\ }_{1,2}$ plane
is just one full hemisphere (traced in the negative orientation):
$\Omega/4\pi=-1/2$. Hence, the constant $C=-1$, leading to the induced
current
\begin{subequations}
\begin{equation}
J^{\nu}=
-
\frac{1}{8\pi}\;\epsilon^{\nu\alpha\beta}\,\epsilon^{abc}\;
n^{\ }_a\,\partial^{\ }_\alpha n^{\ }_b\,\partial^{\ }_\beta n^{\ }_c
\end{equation}
and the induced charge
\begin{equation}
Q=-
\frac{\Omega}{4\pi}.
\end{equation}
\end{subequations}

\subsection{
Abelian formulation, including an axial flux
           }
\label{sec:Abelian-a5}

Here we shall show that the induced charge in the presence of 
vortices in the off-diagonal masses and in an axial vector gauge
potential is the same as that in a problem with constant 
off-diagonal mass and effective Abelian gauge flux. 
To this end, we make a further unitary transformation
on the Lagrange density~(\ref{eq: cal Lchi})
\begin{subequations}
\begin{equation}
\bar\chi=
\bar\xi
U,
\qquad
\chi=
U^{\dag}
\xi.
\end{equation}
The unitary matrix $U$ 
is generated by the $4\times4$ matrices $\Sigma^{\ }_{a}$ and is fixed 
by demanding that it takes the space-time dependent vector
$\boldsymbol{n}$ in the fixed unit vector $\boldsymbol{N}$,
\begin{equation}
\Big(\Sigma^{\ }_{a} N^{\ }_{a}\Big)=
U(x)\,
\Big(
\Sigma^{\ }_{a}\,
n^{\ }_{a}(x)
\Big)\,
U^{\dag}(x).
\end{equation}
\end{subequations}
It follows that $U$ commutes with $\Gamma^{\nu}$,
but not with $\gamma^{\ }_{5}$. 
With $\boldsymbol{N}=(0,0,1)$, this is achieved by choosing~\cite{footnote:rotation}
\begin{subequations}
\begin{equation}
U(x)=
e^{-\mathrm{i}\frac{\beta(x)}{2}\,\Sigma^{\ }_{3}}\;
e^{+\mathrm{i}\frac{\alpha(x)}{2}\,\Sigma^{\ }_{2}}\;
e^{+\mathrm{i}\frac{ \beta(x)}{2}\,\Sigma^{\ }_{3}}
\label{eq:Rot}
\end{equation}
where the polar angle $\alpha$ 
and the azimuthal angle $\beta$ 
parametrize 
$\boldsymbol{n}$,
\begin{equation}
\boldsymbol{n}=
\left(
\sin \alpha
\cos \beta ,
\sin \alpha
\sin \beta ,
\cos \alpha
\right).
\end{equation}
\end{subequations}
The new Lagrange density reads
\begin{subequations}
\begin{equation}
\mathcal{L}^{\ }_{\xi}=
\bar\xi
\left[
\Gamma_{\ }^{\nu}
\left(
\mathrm{i}\partial^{\ }_{\nu}
+
B^{\ }_{\nu}
\right)
-
m\Sigma^{\ }_{3}
\right]
\xi
\label{eq: cal Lxi}
\end{equation}
where the matrix $B^{\ }_{\nu}$ is
\begin{equation}
B^{\ }_{\nu}=
U\gamma^{\ }_{5}U^{\dag}\, A^{\ }_{5\nu}
+
U\mathrm{i}\partial^{\ }_{\nu} U^{\dag}.
\label{eq: def B}
\end{equation}
\end{subequations}

Now once the vector $\boldsymbol{N}$ is fixed to $(0,0,1)$,
all the information about the original mass vortex
and axial vector-gauge vortex is combined in $B^{\ }_{\nu}$.
The induced charge we want to compute is linear
in these potentials 
(with higher orders suppressed by powers of $m^{-1}$).
Indeed to linear order, the current can only depend
on the component of $B^{\ }_{\nu}$ along the 
$a=3$ direction. To see this, consider a further rotation
around the $a=3$ direction by a constant angle $\delta$,
\begin{equation}
B^{\ }_{\nu}\to
e^{+\mathrm{i}\frac{\delta}{2}\Sigma^{\ }_{3}}\,
B^{\ }_{\nu}\,
e^{-\mathrm{i}\frac{\delta}{2}\Sigma^{\ }_{3}}.
\label{eq: rot about N}
\end{equation}
The current is invariant under this rotation,
but the components of $B^{\ }_{\nu}$ along the $a=1,2$
directions do change. Hence, the induced current,
at linear order, must not be a function of these components
and it must depend solely on the component along the $a=3$ 
direction
\begin{eqnarray}
b^{\ }_{\nu}&=&
\frac{1}{4}\mathrm{tr}\,
\left(
\Sigma^{\ }_{3}\;
B^{\ }_{\nu}
\right)
\nonumber\\
&=&
\frac{1}{4}\mathrm{tr}\,
\left[{\Sigma_3 }(
U\gamma^{\ }_{5}U^{\dag}\, A^{\ }_{5\nu}
+
U\mathrm{i}\partial^{\ }_{\nu} U^{\dag})
\right]
\nonumber\\
&=&
\frac{1}{2}\partial^{\ }_{\nu} \beta
-
\frac{1}{2}
\left(
\partial^{\ }_{\nu} \beta
{+}
2A^{\ }_{5\nu}
\right)
\cos \alpha.
\label{eq:bnu explicit}
\end{eqnarray}
We thus arrive at the result that 
the induced current and charge, computed using
the Lagrangian (\ref{eq: cal Lxi}), 
are the same as those computed using the simpler Lagrange
density
\begin{equation}
\bar{\mathcal{L}}^{\ }_{\xi}=
\bar\xi
\left[
\Gamma_{\ }^{\nu}
\left(
\mathrm{i}\partial^{\ }_{\nu}
+
b_{\nu}\,\Sigma^{\ }_{3}
\right)
-
m\Sigma^{\ }_{3}
\right]
\xi.
\label{eq: def tilde cal L chi a}
\end{equation}

Finally, one last change of variables 
\begin{subequations}
\begin{equation}
\bar\xi=
\bar\eta
\;\Sigma^{\ }_{3},
\qquad
\xi=
\eta,
\end{equation}
which does not affect the current, (again because of a trivial Jacobian in the path integral)
and a redefinition of Dirac matrices
\begin{equation}
\bar{\gamma}^{\ }_{\nu}= \Sigma^{\ }_{3}\Gamma^{\ }_{\nu},
\end{equation}
\end{subequations}
which preserves their Clifford algebra,
gives the result that the induced current 
\begin{subequations}
\begin{equation}
\begin{split}
J^{\nu}=&\,
\frac{
\int\mathrm{D}[\bar\eta,\eta]\
e^{
\mathrm{i}
\int d^{3}x\,
\bar{\mathcal{L}}^{\ }_{\eta}
  }\,
\big(
\bar\eta
\bar{\gamma}^{\nu}
\eta
\big)
     }
     {
\int\mathrm{D}[\bar\eta,\eta]\
e^{
\mathrm{i}
\int d^{3}x\,
\bar{\mathcal{L}}^{\ }_{\eta}
  }\,
\hphantom{
\big(
\bar\eta
\bar{\gamma}^{\nu}
\eta
\big)
         }
     }
\end{split}
\end{equation}
can be simply obtained from 
the Lagrange density with the
gauge potential $b^{\ }_{\nu}$ and constant mass $m$
\begin{equation}
\bar{\mathcal{L}}^{\ }_{\eta}=
\bar\eta
\big[
\bar{\gamma}^{\nu}
\left(
\mathrm{i}\partial^{\ }_{\nu}
+
\gamma^{\ }_5 b^{\ }_{\nu}
\right)
-
m
\big]
\eta.
\label{eq: def cal L eta}
\end{equation}
\end{subequations}

The flux due to $b^{\ }_{\nu}$ is the only quantity left 
that retains any information on the mass and axial vortices, and thus
it is the only variable controlling the value of the induced current
and charge. The induced current must be an axial gauge invariant quantity
and thus must be constructed from the axial flux due to
$b^{\ }_{\nu}$. The total charge, in particular,
must be proportional to the total flux 
\begin{equation}
\Phi^{\ }_5=
\frac{1}{2\pi}
\int d^2r \;(\boldsymbol{\partial}\wedge\boldsymbol{b})(\boldsymbol{r}),
\end{equation}
i.e.,
\begin{equation}
Q=
C\;\Phi^{\ }_5 .
\end{equation}
The prefactor $C$ is determined for the special case
without a staggered chemical potential ($\mu\to 0^{+}$)
and without an axial flux, 
in which case $Q=+1/2$. 
In this situation the polar angle 
$\alpha(\boldsymbol{r})\to\pi/2$, $\cos {\alpha}(\boldsymbol{r})\to0$,
and
$b^{\ }_{\nu}(\boldsymbol{r})\to
\frac{1}{2}\,\partial^{\ }_\nu \beta(\boldsymbol{r})$
as $\boldsymbol{r}\to\infty$, 
so that for an antivortex the axial flux due to $b^{\ }_{\nu}$ 
is simply half the vorticity of the azimuthal angle
$\beta(\boldsymbol{r})$:
$\Phi^{\ }_5=-\frac{1}{2}$. 
This fixes the constant $C=-1$.
We conclude that
\begin{equation}
Q=-\Phi^{\ }_5.
\label{eq: final result}
\end{equation}
Equation~(\ref{eq: final result}) is the expression that we seek for
the charge induced by static vortices in the mass and in the axial
vector-gauge fields.~\cite{footnote-ripples} 
We now consider the following two cases (when $\mu>0$).
\begin{itemize}
\item[(i)]
\textit{Static case with staggered chemical potential, no axial flux,
and vorticity $n$ in the mass, i.e.,
$\cos \alpha (\boldsymbol{r})\to\mu/m$ 
as $\boldsymbol{r}\to\infty$ and $a^{\ }_5=0$:}
The flux is 
$\Phi^{\ }_5=n\,\frac{1}{2} [1-\cos \alpha(\boldsymbol{r}\to\infty)]$
and the induced charge equals 
\begin{equation}
Q=-\frac{n}{2}\left(1-\frac{\mu}{m}\right)\;.
\end{equation}
\item[(ii)]
\textit{Static case with staggered chemical potential and an axial vortex that screens
the mass vortex, i.e., 
$\cos \alpha(\boldsymbol{r})\to\mu/m$
and
$\partial^{\ }_\nu \beta(\boldsymbol{r})+2\, A^{\ }_{5\nu}(\boldsymbol{r})\to0$
as $\boldsymbol{r}\to\infty$:}
The last term in Eq.~(\ref{eq:bnu explicit}) drops out and so does
the dependence on the polar angle $\alpha(\boldsymbol{r})$, 
along with the dependence on the staggered chemical potential. 
The flux due to $b^{\ }_{\nu}$ is
simply half the vorticity of the azimuthal angle
$\beta(\boldsymbol{r})$, $\Phi^{\ }_5=\frac{1}{2}\,n$, 
and thus the charge is pinned at the rational value
\begin{equation}
Q=-\frac{n}{2}\;.
\end{equation}
\end{itemize}

\section{
Summary
        }
\label{sec:summary}

The fractional charge induced by vortices supported by
a complex-valued Higgs field carrying a $U(1)$ axial gauge charge of 2
that couples to massive Dirac fermions in $(2+1)$-dimensional space-time was computed by three different techniques based on 
(i) the computation of a spectral asymmetry,
(ii) a one-loop perturbative computation of the conserved fermion-number current,
(iii) expressing the fractional charge in terms of an Abelian axial flux,
respectively.
The fractional charge can vary continuously as a function
of an energy-reflection symmetry breaking parameter and thus can take irrational values. Remarkably, this fractional charge
re-rationalizes to the value $1/2$ taken in the presence
of the spectral energy-reflection symmetry if an axial gauge field
couples covariantly to both the Higgs and Dirac fields.

\section*{
Acknowledgments
         }

This work was supported in part by Grants NSF DMR-0305482 (C.~C. and
C-Y.~H.), DOE DE-FG02-05ER41360 (R.~J.) and DOE DE-FG02-91ER40676
(S-Y.~P.). G.~S. acknowledges the kind hospitality of the Isaac Newton
Institute, Cambridge, and the financial support of NSERC of
Canada.


\begin{thebibliography}{99}
\bibitem{footnote-on-spectral-sym} That the reference states do
possess the energy-reflection symmetry in the case of Hamiltonian
Eq.~(\ref{eq:Hamiltonian}) is argued as follows. In the reference
state, all scalar fields ($\varphi^{\ }_1, \varphi^{\ }_2$, and $\mu$) are
space-time independent, and the axial gauge potential
$\boldsymbol{A}^{\ }_{5} $ is absent. In this case, one can perform a
chiral gauge transformation so as to gauge the phase of the scalar
field and make it purely imaginary (i.e., $\varphi^{\ }_1=0$). The
corresponding Hamiltonian now satisfy $\{\beta,H\}=0$, and thus
eigenvalues of positive and negative energy can be easily related to
one another. An argument can be also constructed for the
energy-reflection symmetry on the reference state on a discrete
lattice model.
\bibitem{footnote:rotation} Whether one should rotate to
$\boldsymbol{N}=(0,0,1)$ or $\boldsymbol{N}=(0,0,-1)$ is determined by
the sign of $\mu$. We choose to work with $\mu>0$ in the computation
in Sec.~\ref{sec:Abelian-a5}.
\bibitem{footnote-ripples} Notice that our expression 
Eq.~(\ref{eq: final result}) can also be used to tie, in a very 
simple way, the local charge
accumulation on a graphene sheet due to ripples, which induce an axial gauge 
field.
\end{thebibliography}
\end{document}